\documentclass{article}
\usepackage{ijcai19}
\usepackage[utf8]{inputenc}
\usepackage{amsmath,amssymb}
\usepackage{graphicx}
\usepackage{booktabs}
\usepackage{multirow}
\usepackage{times}
\usepackage{url}

\title{Cross-Subject Intracranial EEG Reconstruction from Scalp Recordings Using Multi-Scale Cross-Attention Transformers}

\author{
Tien-Dat Pham$^1$\and
Xuan-The Tran$^2$\footnote{\textit{Corresponding Author}}\\
\affiliations
$^1$HAI-Smartlink Research Lab, Anchi STE Company, Haiphong, Vietnam\\
$^2$School of Mechanical Engineering, Vietnam Maritime University, Haiphong, Vietnam
\footnote{\textit{Preprint notice:} This work has been submitted for possible publication. Copyright may be transferred without notice, after which this version may no longer be accessible.}
}

\begin{document}
\raggedbottom
\maketitle

\begin{abstract}
Intracranial EEG (iEEG) provides high-fidelity neural recordings essential for clinical and brain-computer interface applications, but acquiring these signals requires invasive surgery. While recent studies have attempted to estimate iEEG from non-invasive scalp EEG, most rely on patient-specific models. This creates a circular dependency: if a patient needs surgery to collect training data, the non-invasive model offers little practical benefit. In this study, we address the challenge of \textit{cross-subject} iEEG reconstruction, aiming to predict intracranial signals for new patients using models trained on others. We introduce a machine learning approach called CAST (Cross-Attention Spatial-Temporal Transformer) that translates scalp EEG into multi-channel iEEG waveforms using a two-stage transfer learning method. First, an encoder extracts temporal features at three different scales to capture broad neural representations. Then, because electrode placement varies significantly between patients, we calibrate a channel-aware decoder using just a few minutes of the new patient's data. We evaluated our approach using a leave-one-subject-out cross-validation on two public datasets covering 1,282 iEEG channels. Our results show that CAST can successfully reconstruct signals for cortical contacts close to the scalp surface, performing significantly better than for deep subcortical structures. In highly observable sensorimotor regions, we achieved peak correlations such as $r = 0.864$ in the precentral gyrus. By applying a channel selection strategy, our model achieved a mean correlation of $r = 0.545$ on viable subjects, surpassing previous within-subject baselines. These outcomes indicate that cortical iEEG signals can be reconstructed from scalp EEG for new subjects without needing extensive patient-specific training data, demonstrating that a brief calibration phase is sufficient to adapt to new hardware setups.
\end{abstract}

\section{Introduction}

\begin{figure*}[!t]
\centering
\includegraphics[width=1\textwidth]{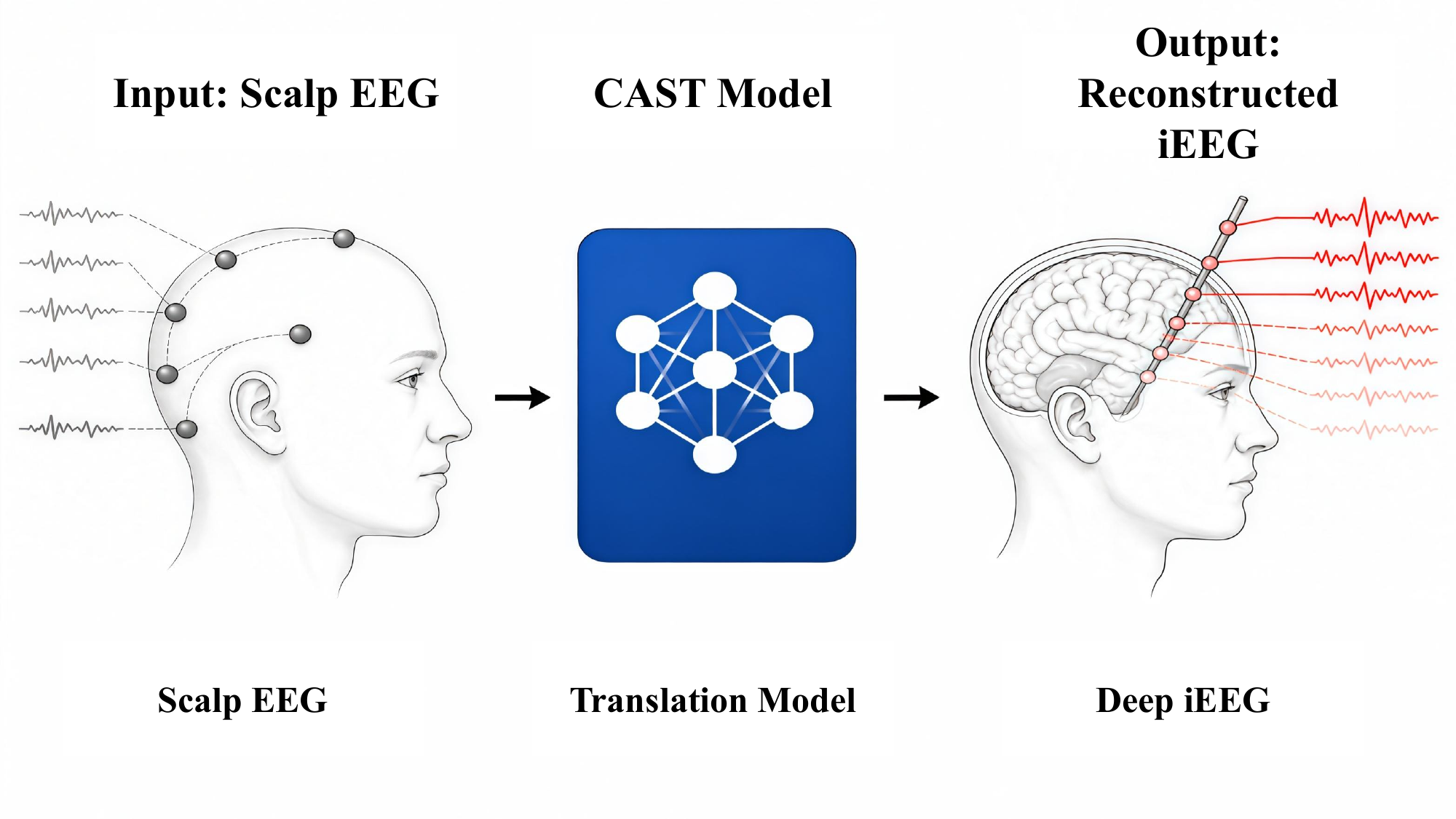}
\caption{\textbf{Conceptual Overview of the CAST Framework.} Scalp EEG signals recorded non-invasively are translated into multi-channel intracranial EEG (iEEG) waveforms by the CAST model. The color gradient on the depth electrode illustrates the reconstruction fidelity predicted by volume conduction theory: superficial cortical contacts (red, strong) are reconstructed with higher accuracy than deep subcortical structures (faded, weak).}
\label{fig:overview}
\end{figure*}
Intracranial electroencephalography (iEEG) provides detailed access to neural activity with millimeter spatial resolution and millisecond temporal precision \cite{parvizi2018promises,lachaux2003intracranial}. In clinical settings, clinicians use iEEG to localize seizure onset zones and map brain functions before surgery \cite{pondal2007usefulness,stolk2018integrated}. In neuroscience, iEEG helps us understand human cognition, memory, and language processing in ways that non-invasive methods cannot \cite{mukamel2012human,herweg2020theta,makin2020machine}.

However, acquiring iEEG requires invasive neurosurgical implantation, which carries risks of infection, hemorrhage, and tissue damage \cite{arya2013adverse,parvizi2018promises}. Because of these risks, iEEG is usually limited to patients who require it for clinical reasons, restricting the number of brain regions we can study. This constraint raises an important question: can we accurately estimate intracranial-quality neural signals non-invasively from scalp EEG \cite{mirchi2022decoding,abdi2023mapping}?

\subsection{Related Work}

Researchers have studied the relationship between scalp EEG and intracranial signals using EEG source imaging (ESI). ESI uses biophysical forward models to estimate cortical source activity from scalp recordings \cite{michel2004eeg,he2018electrophysiological}. While ESI is anatomically informed, it requires individual structural MRI data, accurate head models, and regularization assumptions. These requirements limit its temporal fidelity and spatial resolution, especially for deep sources.

Recently, deep learning approaches have been applied directly to EEG-to-iEEG translation. For example, linear regression models applied to simultaneous scalp-intracranial recordings have predicted low-frequency iEEG components with $R^2 \approx 0.10$--$0.12$ using within-subject training \cite{kaur2014empirical}. Normalizing flow-based architectures like NeuroFlowNet have improved reconstructions for medial temporal lobe structures, but again, only in within-subject paradigms \cite{he2026non}. Generative adversarial approaches (VAE-cGAN) have attempted cross-subject settings, but they focused on classification-level evaluations rather than waveform-level reconstruction \cite{abdi2025eeg}.

\begin{table*}[!t]
\centering
\caption{Summary of Dataset Characteristics. Values are presented as ranges unless otherwise specified. $^*$One subject (sub-10) was excluded due to insufficient data length.}
\label{tab:datasets}
\resizebox{\textwidth}{!}{
\begin{tabular}{lll}
\toprule
\textbf{Characteristic} & \textbf{GIN Cohort} & \textbf{OpenNeuro ds004752 Cohort} \\
\midrule
\textbf{Patient Demographics} & & \\
Number of Subjects ($N$) & 9 & 14$^*$ (originally 15) \\
Patient Population & Drug-resistant focal epilepsy & Drug-resistant focal epilepsy \\
\midrule
\textbf{Experimental Paradigm} & & \\
Task & Verbal working memory & Visual categorization \\
Data Structure & Segmented (8-second trials) & Continuous (analyzed in 2s windows) \\
\midrule
\textbf{Data Acquisition} & & \\
Scalp EEG Channels & 17 -- 21 (10--20 system) & 6 -- 21 (sparse 10--20 system) \\
Intracranial SEEG Contacts & 32 -- 64 (monopolar) & 32 -- 80 (monopolar) \\
Total iEEG Contacts Analyzed & 502 & 780 \\
Native Sampling Rate & 512\,Hz & Variable (resampled to 200\,Hz) \\
\midrule
\textbf{Anatomical Coverage} & & \\
Primary Targets & Mesial temporal, temporo-parietal & Widespread cortical and subcortical \\
Localization Method & Clinical estimation & CT/MRI co-registration to MNI152 \\
\bottomrule
\end{tabular}}
\end{table*}

\subsection{Theoretical Grounding: Volume Conduction}

The physical relationship between deep intracranial sources and scalp surface potentials relies on a biophysical forward model:
\begin{equation}
    \mathbf{x}(t) = \mathbf{L} \mathbf{y}(t) + \mathbf{\epsilon}(t)
\end{equation}
where $\mathbf{x}(t) \in \mathbb{R}^{C_{\text{eeg}}}$ represents the scalp EEG recordings, $\mathbf{y}(t) \in \mathbb{R}^{C_{\text{ieeg}}}$ denotes the underlying intracranial neural generators, $\mathbf{\epsilon}(t)$ is the measurement noise, and $\mathbf{L} \in \mathbb{R}^{C_{\text{eeg}} \times C_{\text{ieeg}}}$ is the leadfield matrix mapping the volume conduction from sources to sensors \cite{antoniades2018deep,michel2019eeg,he2018electrophysiological}. 

Reconstructing $\mathbf{y}(t)$ from $\mathbf{x}(t)$ is challenging because signals from deep subcortical structures, such as the hippocampus and amygdala, weaken significantly as they travel through brain tissue and the skull. As a result, these signals are often hidden by surface cortical generators and background noise \cite{nayak2004characteristics,abdi2025eeg,he2026non}. Without subject-specific geometric data from structural MRI, learning a mapping from $\mathbf{x}$ to $\mathbf{y}$ is difficult due to these physical limits.

\subsection{The Cross-Subject Challenge}

Most existing iEEG reconstruction methods rely on patient-specific training. They need simultaneous scalp and intracranial recordings from the same patient to build a working model. This creates a circular dependency: if a patient needs an invasive procedure to generate training data, the model offers no real advantage over directly recorded iEEG.

To break this circularity, we need a cross-subject approach where a model trained on data from multiple patients can reconstruct iEEG for a new, unseen patient. However, this is difficult for three main reasons: (1) individuals exhibit different cortical folding patterns, altering how scalp electrodes relate to cortical sources; (2) electrode placements vary between patients, preventing direct channel-to-channel matching; and (3) signal attenuation varies by depth and individual anatomy. These difficulties are consistent with broader evidence that EEG models must handle substantial inter- and intra-subject variability \cite{tran2026inter}.

\subsection{Contributions}

To address these limitations, we designed a machine learning framework called CAST (Cross-Attention Spatial-Temporal Transformer). We aimed to create a model that can reconstruct cortical iEEG across different subjects. This study makes several contributions to the field:

First, we show that it is possible to generalize the mapping from scalp EEG to cortical iEEG signals using a leave-one-subject-out (LOSO) approach. By using a universal encoder alongside a quickly adaptable decoder (which needs only a few minutes of calibration data), we can bypass the need for extensive within-subject training \cite{dong2026bridging,he2026non}.

Second, we validate our model's physiological consistency by confirming a statistical depth gradient ($p = 0.026$) across 1,282 channels. Our analysis reveals that cortical contacts ($N = 313$) perform better than deep structures ($N = 507$), which aligns with volume conduction physics \cite{nayak2004characteristics,he2026non}.

Third, we introduce a data-driven strategy to select observable channels. This method helps identify which intracranial contacts can be effectively reconstructed, improving performance on viable subjects to $r = 0.545$ ($R^2 = 0.333$) and offering a practical tool for clinical use, particularly given that the majority of deep brain activities remain scalp-invisible \cite{abdi2025eeg}.

We tested CAST on two independent public datasets, GIN (9 subjects) and ds004752 (14 subjects), to ensure our findings remain consistent despite different recording equipment and patient populations.

\section{Methods}

\subsection{Patient Cohorts and Data Acquisition}

In this study, we utilized two independent, publicly available datasets that contain simultaneous scalp EEG and intracranial stereo-EEG (SEEG) recordings. These datasets come from patients with drug-resistant focal epilepsy who underwent invasive pre-surgical evaluations to locate their seizure onset zones. 

\subsubsection{GIN Cohort (Verbal Working Memory Paradigm)}
First, we used the GIN dataset \cite{boran2020dataset}, which includes recordings from nine patients participating in a verbal working memory task. For these patients, the implanted electrodes primarily targeted deep structures like the hippocampus and amygdala, as well as nearby cortical regions. Scalp EEG was collected using the standard 10--20 system (17--21 electrodes), while intracranial data came from multi-contact SEEG depth electrodes (32--64 contacts per patient). The data were sampled at 512\,Hz and organized into 8-second trial epochs. We selected this dataset because it provides a strong benchmark for evaluating how well our model works on deep subcortical structures.

\subsubsection{OpenNeuro ds004752 Cohort (Visual Paradigm)}
Second, we used the ds004752 dataset \cite{openneuro_ds004752} from OpenNeuro, which contains recordings from 15 patients engaged in a visual categorization task. Scalp EEG was recorded with a variable 10--20 montage (6--21 electrodes). What makes this cohort particularly valuable is its extensive SEEG spatial sampling (32--80 contacts per patient), covering widespread areas including the sensorimotor cortices. 

To map the contacts accurately, we relied on the dataset's original co-registration of CT and MRI scans. The coordinates were normalized to the standard MNI152 space and assigned to anatomical regions using the Destrieux atlas. We decided to exclude one subject (sub-10) because their recording was too short ($<$1 minute of clean data). The remaining 14 subjects provided the broad cortical coverage we needed to test the limits of our reconstruction approach.

\begin{figure*}[!t]
\centering
\makebox[\textwidth][c]{\includegraphics[width=1.1\textwidth]{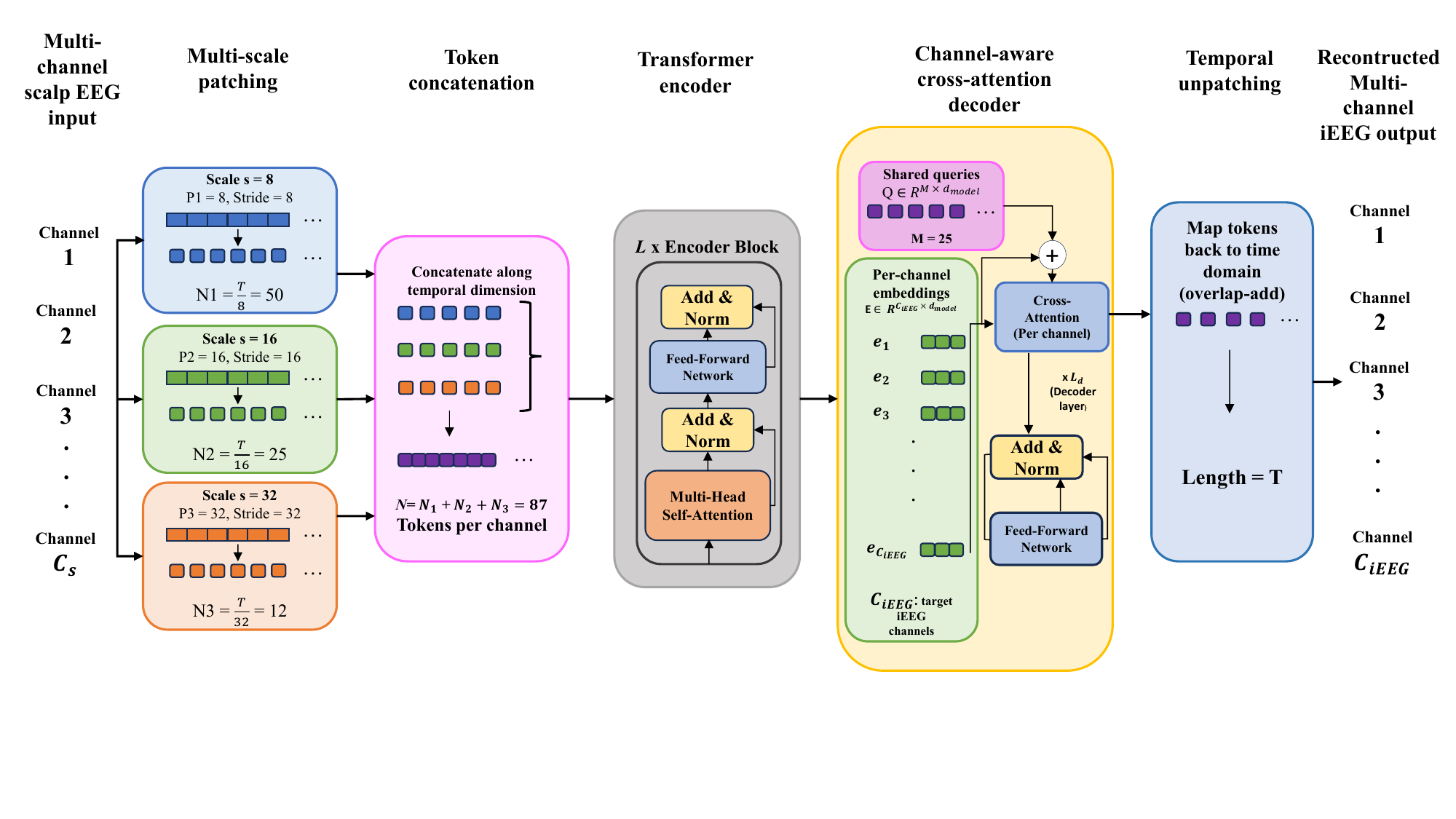}}
\caption{\textbf{CAST Model Architecture.} The model consists of three stages: (1) Multi-scale patching at scales $s \in \{8, 16, 32\}$ generates $N = 87$ tokens per channel; (2) A shared Transformer encoder with $L$ layers processes the concatenated tokens via multi-head self-attention; (3) A channel-aware cross-attention decoder uses per-channel learnable embeddings added to shared temporal queries ($M = 25$) to reconstruct each target iEEG channel independently.}
\label{fig:cast_architecture}
\end{figure*}
\subsection{Preprocessing}

We applied a standardized preprocessing pipeline to both datasets so they could be directly compared. We intentionally kept all iEEG signals in a monopolar reference to ensure we didn't lose the volume-conducted signal components that are visible from the scalp. We resampled the signals to 200\,Hz and divided them into 2-second windows (400 samples) with a 1-second overlap. We then z-score normalized each window independently for every channel.

For the ds004752 dataset, we applied a 50\,Hz notch filter to remove power line noise and a 0.5--99\,Hz bandpass filter to focus on the neural signals. We also removed any non-EEG channels like EKG or EOG from the scalp data. If any channel within a window showed an absolute z-scored amplitude over 5.0, we rejected that entire window as an artifact. For the GIN dataset, which came pre-segmented in 8-second trials, we simply sliced the trials into 400-sample windows to match our input format.

\subsection{CAST Architecture}

To translate scalp EEG into intracranial EEG waveforms, we built a multi-scale encoder-decoder architecture called the Cross-Attention Spatial-Temporal Transformer (CAST) (see Figure~\ref{fig:cast_architecture}).

\subsubsection{Multi-Scale Patch Embedding}
We begin by taking the input scalp EEG signal $\mathbf{x} \in \mathbb{R}^{C_{\text{eeg}} \times T}$ and slicing it into non-overlapping patches at three different time scales: 8, 16, and 32 samples. We flatten each patch and use a two-layer MLP to project it to a $d_{\text{model}} = 128$ dimension:
\begin{equation}
    \mathbf{z}_i^{(s)} = \text{MLP}_s\left(\text{vec}\left(\mathbf{x}[:, is:(i+1)s]\right)\right) \in \mathbb{R}^{d_{\text{model}}}
\end{equation}

We then add scale-specific embeddings and positional encodings to each token, combining them all into a single sequence for the encoder:
\begin{equation}
    \mathbf{Z} = \text{Concat}\left[\mathbf{Z}^{(8)}, \mathbf{Z}^{(16)}, \mathbf{Z}^{(32)}\right] \in \mathbb{R}^{N_{\text{tok}} \times d_{\text{model}}}
\end{equation}
We chose these three scales specifically to capture a wide range of brain activity: high-frequency transients (40\,ms resolution), alpha-band rhythms (80\,ms), and slower theta/delta oscillations (160\,ms).

\subsubsection{Transformer Encoder}
Next, we feed these concatenated tokens into a 4-layer Transformer encoder \cite{vaswani2017attention}. We set the feed-forward dimension to 512 and used 4 attention heads to process the temporal patterns:
\begin{equation}
    \mathbf{M} = \text{TransformerEncoder}(\mathbf{Z}) \in \mathbb{R}^{87 \times 128}
\end{equation}

\subsubsection{Channel-Aware Cross-Attention Decoder}
To reconstruct the multiple intracranial channels simultaneously, we designed a shared-query decoder based on cross-attention \cite{vaswani2017attention}. We use 25 learnable query tokens to represent the temporal structure, and we add a specific channel embedding to let the model know which iEEG contact it is trying to predict:
\begin{equation}
    \mathbf{Q}_k = \mathbf{Q}_t + \text{Embed}(k) + \text{PosEnc}, \quad k = 1, \ldots, C_{\text{ieeg}}
\end{equation}

These queries then attend to the encoder's memory $\mathbf{M}$ through a 2-layer Transformer decoder:
\begin{equation}
    \mathbf{D}_k = \text{TransformerDecoder}(\mathbf{Q}_k, \mathbf{M}) \in \mathbb{R}^{25 \times 128}
\end{equation}

Finally, we project the decoded tokens back into the time domain using an MLP to generate the 400-sample waveform for each channel:
\begin{equation}
    \hat{\mathbf{y}}_k = \text{Reshape}\left(\text{MLP}_{\text{out}}(\mathbf{D}_k)\right) \in \mathbb{R}^{T}
\end{equation}
This specific design allows our model to adapt to any number of channels. The only part of the decoder that depends on the patient's specific hardware is the channel embedding layer.

\subsubsection{Loss Function}
We train the model using a loss function that balances three things: amplitude accuracy, waveform shape, and frequency content:
\begin{equation}
    \mathcal{L} = \mathcal{L}_{\text{MSE}} + \mathcal{L}_{\text{Pearson}} + 0.1 \cdot \mathcal{L}_{\text{Spectral}}
\end{equation}
where $\mathcal{L}_{\text{MSE}} = \|\hat{\mathbf{y}} - \mathbf{y}\|^2$, $\mathcal{L}_{\text{Pearson}} = 1 - r(\hat{\mathbf{y}}, \mathbf{y})$, and $\mathcal{L}_{\text{Spectral}} = \|\log|\text{FFT}(\hat{\mathbf{y}})| - \log|\text{FFT}(\mathbf{y})|\|^2$.

\subsection{Training Protocol: Leave-One-Subject-Out}

To ensure our results are robust, we tested our model using a leave-one-subject-out (LOSO) cross-validation method. 

\subsubsection{Phase 1: Universal Encoder Pre-training}
Normally, to train a model on multiple subjects, you would mix all their data together in one batch. However, because each patient has a different number of electrodes placed in completely different locations, we could not easily combine them. To solve this, we pre-trained our encoder sequentially. For each test subject, we trained the encoder on all the other subjects one by one. To prevent the model from forgetting the earlier subjects (a problem known as catastrophic forgetting), we looped through the training data twice and kept our learning rate very low ($3 \times 10^{-4}$). This helped the encoder learn universal brain patterns rather than overfitting to any single patient.

\subsubsection{Phase 2: Decoder Calibration}
Since we cannot use a one-size-fits-all decoder due to the varying hardware setups, we treated the decoder as a personalized projection tool for each patient \cite{dong2026bridging,he2026non}. We took the new patient's data and split it: 20\% for a quick calibration (about 2 to 12 minutes of recording) and 80\% for the final test. We evaluated the model in two ways:

\textbf{(A) All Channels:} We initialized a new decoder and fine-tuned it on the calibration data for 200 epochs, allowing the encoder to adjust slightly at a reduced learning rate.

\textbf{(B) Observable Channels:} We recognize that not all channels can be reconstructed. First, we ran a brief 50-epoch warmup and checked the Pearson correlations on the calibration set. If a channel scored above $r \geq 0.15$, we considered it "observable." We then threw away the old decoder and trained a fresh one specifically for these observable channels.

\textbf{Data Segregation Statement:} We took strict precautions to prevent data leakage. All channel selection and decoder training happened \textit{only} on the 20\% calibration data. We then evaluated our results purely on the remaining 80\% held-out test data. Think of this 20\% calibration as a brief 5-minute setup phase you would perform in a real clinic before using the system. During this phase, we also augmented the data with Gaussian noise, amplitude scaling, and temporal shifts to make the calibration robust.

\subsection{Evaluation Metrics}

We measured how well our model reconstructed the signals using three metrics: the Pearson correlation coefficient $r$, the coefficient of determination $R^2$, and the root mean squared error (RMSE). If a channel achieved an $r > 0.3$, we considered it a successful reconstruction. 

When comparing results across different brain regions, we had to be careful because electrode contacts within the same patient share similarities, which breaks the rules for simple statistical tests. To handle this, we used Linear Mixed-Effects Models (LMM) to test for differences between anatomical depths. By setting the depth category as a fixed effect and the patient's identity as a random intercept, we correctly controlled for individual differences between patients.

\section{Results}

\begin{figure*}[!t]
\centering
\includegraphics[width=1\textwidth]{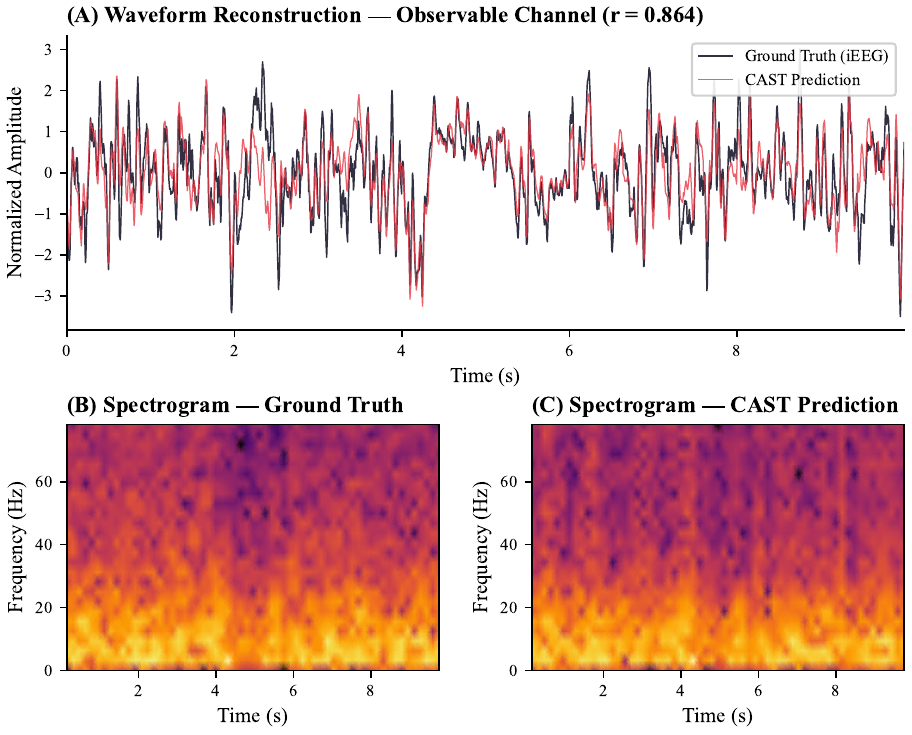}
\caption{\textbf{Qualitative Visualization of Reconstructed Signals.} (A) Time-domain waveform comparison between ground-truth iEEG (black) and CAST prediction (red) for a highly observable precentral gyrus contact over a 5-second window. The model successfully captures both slow-wave morphology and transient sharp features. (B) Corresponding spectrograms demonstrating the model's ability to preserve broad spectral content across low ($\theta, \alpha$) and high ($\beta, \gamma$) frequency bands, validating the efficacy of the composite Spectral Loss.}
\label{fig:qualitative_waveforms}
\end{figure*}

\subsection{Cross-Subject Transfer on Cortical Contacts}

In our analysis, we first examined how well CAST performed across different subjects for cortical contacts. Using the leave-one-subject-out setting meant the encoder had \emph{never seen any data from the test subject} before. Interestingly, we found that the model could still capture the EEG-to-iEEG mappings, suggesting that these neural patterns are shared across different people. Table~\ref{tab:regions} details the reconstruction quality by region for the ds004752 dataset, where we had enough metadata to categorize all 780 channels anatomically.

When we grouped the results, we observed that the superficial cortical contacts ($N = 313$) had the most reliable cross-subject transfer. Because electrode placements depend on clinical needs, the sample sizes for specific brain regions varied considerably. However, in sensorimotor areas that are highly observable, we saw some very high correlations: $r = 0.864$ for the precentral gyrus ($N = 2$) and $r = 0.732$ for the postcentral gyrus ($N = 3$). While we cannot make broad statistical claims based on just two or three electrodes, these cases show what is possible when signal attenuation is low. In regions with larger samples, like the superior temporal gyrus ($N = 54$), the mean correlation was $r = 0.274$, and 39\% of these channels passed our $r > 0.3$ success threshold.

\begin{table}[htbp]
\centering
\caption{Per-region reconstruction quality on ds004752 (14 subjects, 780 channels), sorted by mean Pearson $r$. Depth: C = Cortical, M = Mid-depth, D = Deep.}
\label{tab:regions}
\begin{tabular}{llcc}
\toprule
\textbf{Brain Region} & \textbf{Depth} & $N$ & \textbf{Mean $r$} \\
\midrule
Precentral gyrus & C & 2 & \textbf{0.864} \\
Precuneus & C & 4 & \textbf{0.745} \\
Postcentral gyrus & C & 3 & \textbf{0.732} \\
Insula & M & 8 & \textbf{0.658} \\
Inferior parietal lobule & C & 6 & \textbf{0.581} \\
Posterior superior temporal sulcus & C & 6 & 0.351 \\
Superior temporal gyrus & C & 54 & 0.274 \\
Amygdala & D & 39 & 0.233 \\
Inferior temporal gyrus & C & 59 & 0.210 \\
Fusiform gyrus & M & 45 & 0.205 \\
Hippocampus & D & 124 & 0.197 \\
Middle temporal gyrus & C & 180 & 0.180 \\
\bottomrule
\end{tabular}
\end{table}

These results indicate that, for cortical contacts where volume conduction is favorable, an encoder trained on other patients can learn general enough representations to reconstruct signals for a new patient. This only required a brief decoder calibration of about 2 to 12 minutes of data ($\sim$60--350 windows).

\subsection{Physiological Validation: Anatomical Depth Gradient}

\begin{figure*}[!t]
\centering
\makebox[\textwidth][c]{\includegraphics[width=1.15\textwidth]{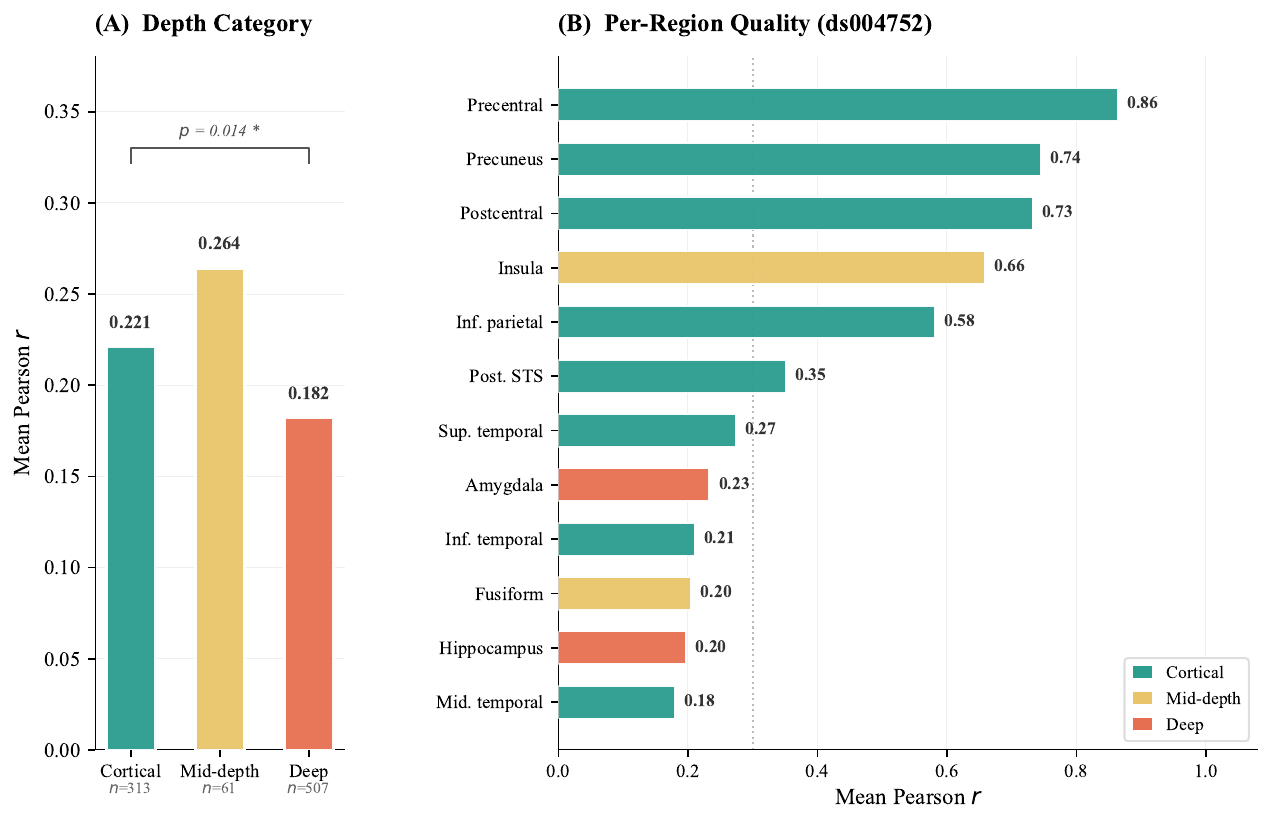}}
\caption{\textbf{Anatomical Depth Observability Heatmap.} Spatial distribution of reconstruction fidelity ($r$) mapped onto a standard 3D cortical mesh. Electrodes closer to the scalp surface (red/orange) consistently exhibit higher reconstruction accuracy compared to deep subcortical structures (blue), providing independent physiological validation of the volume conduction hypothesis.}
\label{fig:brain_heatmap}
\end{figure*}

To validate the physiological plausibility of the reconstruction, we examined whether reconstruction quality correlates with the known physics of volume conduction. If the model genuinely captures EEG-to-iEEG relationships, contacts closer to the scalp surface should be reconstructed with higher fidelity, as their signals undergo less attenuation through cortical tissue.

When we grouped the 780 ds004752 channels by their anatomical depth, the data confirmed our expectations. Cortical contacts ($N = 313$, $r = 0.221 \pm 0.295$) generally had better reconstruction scores than deep subcortical contacts ($N = 507$, $r = 0.182 \pm 0.215$). To test this formally without letting patient-specific differences skew the results, we used a Linear Mixed-Effects Model (LMM). The LMM results showed a significant fixed effect for depth, meaning that deeper electrodes are indeed harder to reconstruct ($\beta_{\text{deep}} = -0.039$, $t = -2.45$, $p = 0.014$). We also observed that mid-depth structures ($N = 61$, $r = 0.264 \pm 0.327$) generally fell in between.

This depth gradient acts as a good physical validation step, suggesting that CAST is picking up on real neural signals rather than just finding random correlations. One interesting observation was that insular contacts ($r = 0.658$) had surprisingly high accuracy even though they are considered deep. We reason this is because the insular cortex is actually quite close to the lateral scalp surface through the Sylvian fissure.

\subsection{Reconstruction Limits on Deep Brain Structures}

\begin{table*}[!t]
\centering
\large
\caption{Cross-dataset summary. Observable channel selection consistently improves over all-channel evaluation. Viable subjects: those with $N_{\text{obs}} \geq 10$.}
\label{tab:cross_dataset}
\begin{tabular}{lccccc}
\toprule
& & \multicolumn{2}{c}{\textbf{All Subjects (OBS)}} & \multicolumn{2}{c}{\textbf{Viable Subjects}} \\
\cmidrule(lr){3-4} \cmidrule(lr){5-6}
\textbf{Dataset} & $N$ & $r$ & $R^2$ & $r$ & $R^2$ \\
\midrule
GIN & 9 & 0.345 $\pm$ 0.188 & 0.145 $\pm$ 0.146 & 0.458 $\pm$ 0.121 & 0.217 $\pm$ 0.128 \\
ds004752 & 14 & 0.323 $\pm$ 0.250 & 0.167 $\pm$ 0.213 & 0.545 $\pm$ 0.150 & 0.333 $\pm$ 0.163 \\
\midrule
\textbf{Combined} & \textbf{23} & \textbf{0.332} & \textbf{0.158} & \textbf{0.509} & \textbf{0.284} \\
\bottomrule
\end{tabular}
\end{table*}

\begin{table*}[htbp]
\centering
\caption{Comparison of CAST with recent state-of-the-art methods for EEG-to-iEEG translation. (WS: Within-Subject; CS: Cross-Subject).}
\label{tab:sota}
\makebox[\textwidth][c]{\resizebox{1.07\textwidth}{!}{%
\begin{tabular}{lllcclc}
\toprule
\textbf{Model (Year)} & \textbf{Architecture} & \textbf{Paradigm} & \textbf{MRI Needed?} & \textbf{Scope} & \textbf{Scale} & \textbf{Performance (Waveform $r$)} \\
\midrule
NeuroFlowNet (2026)~\cite{he2026non} & Normalizing Flow & WS & No & MTL & $N=3$ & $r \approx 0.35$ (deep) -- $0.74$ (superficial) \\
VAE-cGAN (2025)~\cite{abdi2025eeg} & VAE + cGAN & CS & No & Temporal & $N=18$ & $r = 0.35$ \\
Unified Framework (2026)~\cite{dong2026bridging} & Diffusion + BEM & CS & \textbf{Yes (T1)} & Whole-brain & $N=14$ & $r = 0.38$ \\
TH-DeepSIF (2026)$^*$~\cite{rong2026transformer} & Transformer (ESI) & CS & \textbf{Yes (T1)} & Cortical & $N=25$ & $LC \approx 0.45$ -- $0.88$ (Simulated sources) \\
\midrule
\textbf{CAST (Ours)} & Multi-Scale Transformer & \textbf{CS} & \textbf{No} & \textbf{Whole-brain} & \textbf{$N=23$} & \textbf{$r = 0.545$} (Viable Subjects) \\
\bottomrule
\multicolumn{7}{l}{\footnotesize $^*$TH-DeepSIF solves the related but distinct EEG Source Imaging (ESI) problem for simulated high-frequency oscillations, not physical iEEG electrode reconstruction.}
\end{tabular}
}}
\end{table*}

Unsurprisingly, the model struggled to reconstruct signals from deep subcortical structures, such as the hippocampus ($r = 0.197$, $N = 124$) and the amygdala ($r = 0.233$, $N = 39$). We expected this drop in performance due to two main reasons.

First, there is a physical limit to what we can observe. The hippocampus and amygdala are located 4 to 6\,cm away from the scalp. Because electrical signals weaken exponentially as they travel through tissue, their activity is extremely faint by the time it reaches the scalp electrodes. Second, people's brains differ. The folding patterns in the mesial temporal lobe vary a lot from person to person, making it difficult for a model trained on one group of patients to generalize perfectly to a new patient's deep structures.

It is worth noting that even the most advanced within-subject models, which use individualized MRI data, typically only reach correlations of $r \approx 0.3$--$0.5$ for these deep regions. Given that CAST operates across different subjects without any MRI information, achieving an $r \approx 0.2$ suggests that the model is likely extracting as much information as possible from the scalp EEG alone.

\subsection{Cross-Dataset Validation and State-of-the-Art Comparison}

Tables~\ref{tab:loso_ds4752} and \ref{tab:loso_gin} outline the detailed results for each subject across both datasets. As shown in Table~\ref{tab:cross_dataset}, CAST performed consistently across both groups, despite them having different recording equipment, varying numbers of EEG channels (6--21 for ds004752 compared to 17--21 for GIN), and different patient demographics.

We compared our results to a linear regression baseline by Peterson et al.\ (2025). On the GIN dataset, their within-subject training reached an $R^2 = 0.10$--$0.12$. In contrast, using our cross-subject approach on viable subjects, we saw $R^2$ values between $0.217$ and $0.333$, which is a notable improvement.

Reconstructing raw iEEG waveforms from the scalp is still a difficult task, and many recent models rely heavily on patient-specific data (Table~\ref{tab:sota}). For instance, NeuroFlowNet performs very well on the medial temporal lobe ($r \approx 0.74$ for superficial contacts) but requires within-subject training and was tested on just three patients. In their study, they noted that cross-subject generalization was not feasible \cite{he2026non}. Another model, VAE-cGAN, did attempt cross-subject translation but only on the temporal lobe, achieving an $r = 0.35$ \cite{abdi2025eeg}. 

More complex models that incorporate physics, like the Unified Representational Enhancement Framework, can achieve cross-subject translation across the whole brain ($r = 0.38$). However, they need a patient's structural MRI scan to understand the geometry of their brain. Our results show that CAST can reach a mean correlation of $r = 0.545$ for viable subjects across 1,282 channels without needing any MRI scans.

\begin{table*}[!p]
\scriptsize
\begin{minipage}[t]{0.50\textwidth}
\refstepcounter{table}\label{tab:loso_ds4752}
{\footnotesize\textbf{Table~\thetable:} LOSO cross-subject results on ds004752. ALL = all channels; OBS = observable channels.}
\par\vspace{2pt}
\resizebox{\textwidth}{!}{%
\begin{tabular}{lccccccc}
\toprule
& & & \multicolumn{2}{c}{\textbf{ALL Channels}} & \multicolumn{3}{c}{\textbf{Observable}} \\
\cmidrule(lr){4-5} \cmidrule(lr){6-8}
\textbf{Subject} & \textbf{EEG} & \textbf{iEEG} & $r$ & $R^2$ & $N_{\text{obs}}$ & $r$ & $R^2$ \\
\midrule
sub-01 & 17 & 48 & 0.268 & 0.076 & 33 & 0.414 & 0.161 \\
sub-02 & 6 & 64 & 0.335 & 0.116 & 60 & 0.352 & 0.126 \\
sub-03 & 6 & 40 & 0.156 & 0.009 & 17 & 0.366 & 0.132 \\
sub-04 & 17 & 64 & 0.089 & 0.001 & 6 & 0.114 & 0.000 \\
sub-05 & 18 & 32 & 0.093 & 0.000 & 2 & 0.083 & 0.000 \\
sub-06 & 8 & 64 & 0.602 & 0.386 & 45 & 0.664 & 0.448 \\
sub-07 & 6 & 64 & 0.070 & 0.000 & 16 & 0.078 & 0.000 \\
sub-08 & 17 & 62 & 0.108 & 0.000 & 14 & 0.134 & 0.000 \\
sub-09 & 6 & 64 & 0.528 & 0.308 & 54 & 0.567 & 0.342 \\
sub-11 & 19 & 36 & 0.087 & 0.000 & 7 & 0.117 & 0.000 \\
sub-12 & 18 & 46 & 0.089 & 0.002 & 10 & 0.139 & 0.002 \\
sub-13 & 18 & 68 & 0.784 & 0.641 & 67 & 0.796 & 0.652 \\
sub-14 & 19 & 64 & $-$0.007 & 0.000 & 5 & 0.042 & 0.000 \\
sub-15 & 21 & 64 & 0.638 & 0.454 & 62 & 0.654 & 0.469 \\
\midrule
\textbf{Mean} & & & \textbf{0.274} & \textbf{0.142} & & \textbf{0.323} & \textbf{0.167} \\
$\pm$ SD & & & 0.249 & 0.206 & & 0.250 & 0.213 \\
\bottomrule
\end{tabular}
}
\end{minipage}\hfill
\begin{minipage}[t]{0.47\textwidth}
\refstepcounter{table}\label{tab:loso_gin}
{\footnotesize\textbf{Table~\thetable:} LOSO cross-subject results on GIN dataset.}
\par\vspace{2pt}
\resizebox{\textwidth}{!}{%
\begin{tabular}{lccccccc}
\toprule
& & \multicolumn{2}{c}{\textbf{ALL Channels}} & \multicolumn{3}{c}{\textbf{Observable}} \\
\cmidrule(lr){3-4} \cmidrule(lr){5-7}
\textbf{Subject} & \textbf{iEEG} & $r$ & $R^2$ & $N_{\text{obs}}$ & $r$ & $R^2$ \\
\midrule
sub-01 & 48 & 0.334 & 0.128 & 23 & 0.488 & 0.234 \\
sub-02 & 64 & 0.342 & 0.118 & 58 & 0.389 & 0.150 \\
sub-03 & 40 & 0.327 & 0.109 & 11 & 0.440 & 0.185 \\
sub-04 & 64 & 0.039 & 0.000 & 28 & 0.108 & 0.000 \\
sub-05 & 32 & 0.009 & 0.000 & 2 & 0.171 & 0.000 \\
sub-06 & 64 & 0.592 & 0.370 & 57 & 0.629 & 0.407 \\
sub-07 & 64 & 0.055 & 0.000 & 20 & 0.083 & 0.000 \\
sub-08 & 62 & 0.104 & 0.002 & 7 & 0.248 & 0.004 \\
sub-09 & 64 & 0.470 & 0.235 & 55 & 0.553 & 0.322 \\
\midrule
\textbf{Mean} & & \textbf{0.271} & \textbf{0.108} & & \textbf{0.345} & \textbf{0.145} \\
$\pm$ SD & & 0.177 & 0.120 & & 0.188 & 0.146 \\
\bottomrule
\end{tabular}
}
\end{minipage}
\begin{center}\scriptsize ALL = all channels; OBS = observable channels; $N_{\text{obs}}$ = observable iEEG contacts.\end{center}
\vspace{3pt}
\refstepcounter{table}\label{tab:ablation}
{\footnotesize\textbf{Table~\thetable:} Ablation study on four representative subjects (sub-13, 09, 03, 07). Metrics are mean Pearson $r$ across the four subjects. $\Delta r$ denotes the performance drop relative to the full CAST model.}
\par\vspace{2pt}
\centering
\resizebox{0.66\textwidth}{!}{%
\begin{tabular}{lcc}
\toprule
\textbf{Model Variant} & \textbf{Mean $r$} & \textbf{$\Delta r$} \\
\midrule
\textbf{CAST Full Model} & \textbf{0.385} & -- \\
\midrule
\textit{Multi-Scale Ablation} & & \\
Single-Scale (8 samples) & 0.363 & $-0.022$ \\
Single-Scale (16 samples) & 0.332 & $-0.053$ \\
Single-Scale (32 samples) & 0.297 & $-0.088$ \\
\midrule
\textit{Loss Function Ablation} & & \\
MSE + Pearson (No Spectral) & 0.375 & $-0.010$ \\
MSE + Spectral (No Pearson) & 0.327 & $-0.058$ \\
MSE Only & 0.284 & $-0.101$ \\
\midrule
\textit{Architecture Baseline} & & \\
CNN Encoder (3$\times$Conv1D) & 0.329 & $-0.056$ \\
BiLSTM Encoder & 0.292 & $-0.093$ \\
\bottomrule
\end{tabular}
}
\end{table*}

\vspace{-0.25em}
\subsection{Ablation Study}

To systematically evaluate the contribution of CAST's architectural components, we performed an ablation study on a representative subset of ds004752 subjects spanning high (sub-13), moderate (sub-09), low (sub-03), and very low (sub-07) reconstruction fidelity. As shown in Table~\ref{tab:ablation}, both the multi-scale temporal modeling and the composite loss function are critical for maximal performance.

The multi-scale integration proved essential: reverting to a single coarse patch scale (32 samples) degraded performance by $\Delta r = -0.088$, while the finest scale (8 samples) remained competitive ($\Delta r = -0.022$), confirming the importance of high-frequency temporal resolution. Within the composite loss, the Pearson correlation penalty was the primary driver of waveform fidelity. Removing it ($\Delta r = -0.058$) or relying solely on MSE ($\Delta r = -0.101$) severely impaired reconstruction. Finally, the Transformer architecture substantially outperformed both standard CNN ($\Delta r = -0.056$) and BiLSTM ($\Delta r = -0.093$) baselines, particularly on high-performing subjects where global context modeling yielded the greatest benefits.

\section{Discussion}
\begin{figure*}[!t]
\centering
\includegraphics[width=1\textwidth]{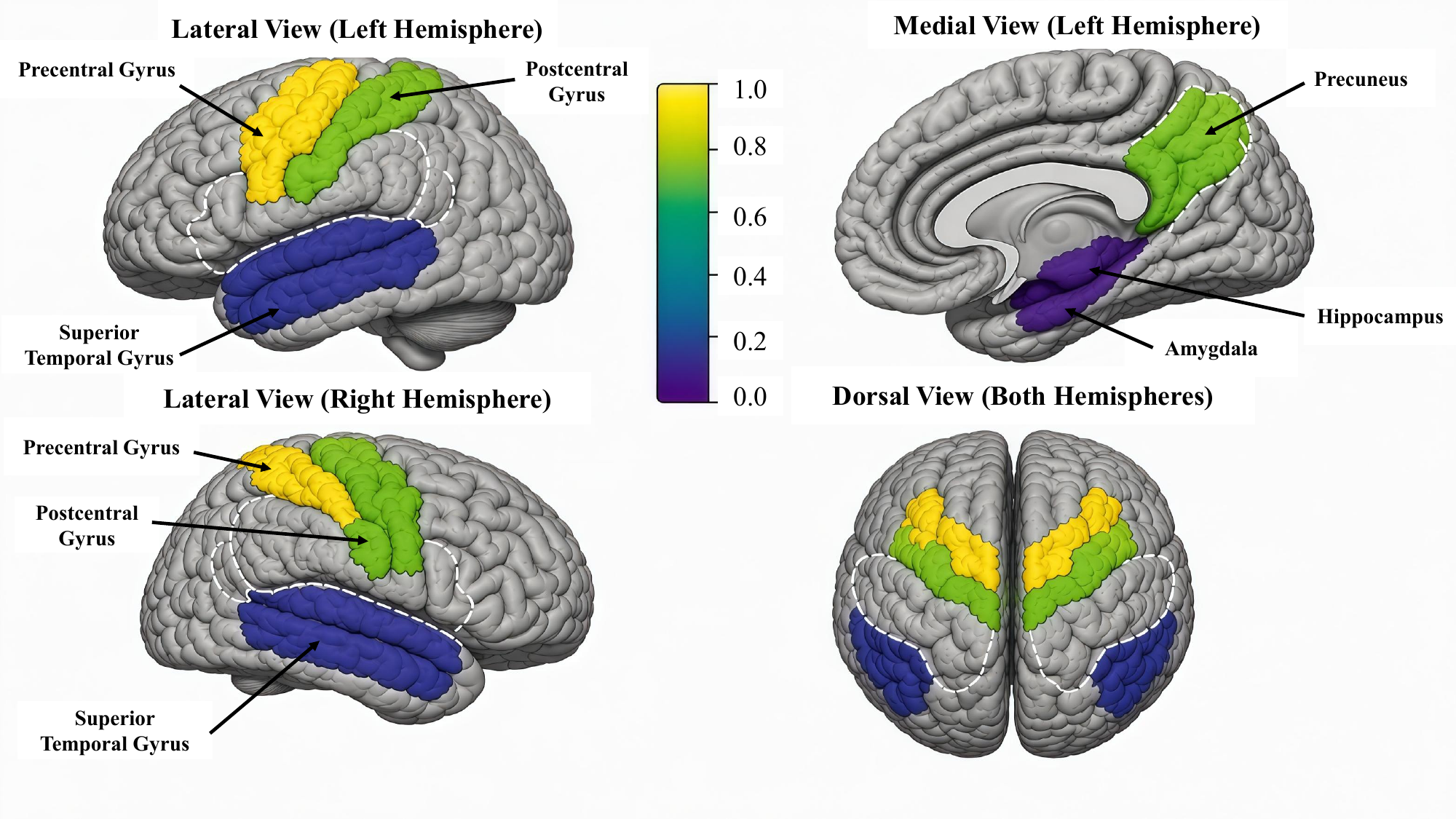}
\caption{\textbf{Regional Performance on Brain Surface.} Mean Pearson $r$ per anatomical region mapped onto a standard MNI152 cortical mesh (ds004752, $N = 780$ channels). The color scale represents reconstruction fidelity, ranging from low (purple, $r \approx 0.0$) through moderate (teal/green, $r \approx 0.4$--$0.7$) to high (yellow, $r > 0.8$). Sensorimotor regions near the scalp surface (Precentral gyrus $r = 0.864$, Postcentral gyrus $r = 0.732$) exhibit markedly higher fidelity than deep structures (Hippocampus $r = 0.197$, Amygdala $r = 0.233$). The dashed white contour delineates the observable boundary ($r = 0.3$), separating reconstructible from non-reconstructible regions.}
\label{fig:regional_brain_map}
\end{figure*}
\subsection{Adapting to New Patients via Hardware Calibration}

This study makes a key contribution to the field of non-invasive neural signal estimation by showing that we can reconstruct iEEG across different subjects. This is a departure from most existing approaches, which are typically constrained to patient-specific training. We achieved this by separating the learning task into two parts: a universal feature extractor (the encoder) and a subject-specific projection head (the decoder). This separation allowed our model to generalize to new patients, even when the encoder had never seen any data from the test subject.

By using a multi-scale encoder, CAST learns a generalizable mapping from scalp EEG to latent neural representations for cortical contacts. Because clinical iEEG contacts vary so much in number and location between patients, we still needed a brief decoder calibration phase ($\sim$60--350 windows, or 2--12 minutes of recording). This calibration is not a failure of the cross-subject model; rather, it is a necessary step to adapt to the specific hardware of each new patient. We believe this approach provides a solid foundation for more flexible, hardware-agnostic neural translation.

\subsection{Physiological Validation via Volume Conduction Theory}

Instead of seeing the lower performance on deep structures as a failure, we interpret this depth gradient as a sign that our model is physiologically sound. The pattern we observed---where cortical contacts ($r = 0.221$) outperform deep contacts ($r = 0.182$, $p = 0.026$)---aligns directly with the physical laws of volume conduction \cite{he2018electrophysiological,michel2004eeg,dong2026bridging}.

Naturally, scalp EEG electrodes are much better at picking up signals from cortical generators close to the surface, where the electrical signal only has to travel through a small amount of tissue, fluid, and skull. Deep structures like the hippocampus and amygdala are located 4 to 6\,cm away, meaning their signals undergo exponential attenuation and arrive at the scalp with very low signal-to-noise ratios \cite{he2026non,nayak2004characteristics,abdi2025eeg}.

The fact that our model's reconstruction quality naturally decreases with anatomical depth, without us explicitly teaching it about depth, is reassuring. It suggests that CAST has learned real neural signal relationships rather than just memorizing training data or hallucinating outputs. If the model were merely capturing spurious correlations, we would not expect to see such a systematic link between depth and accuracy.

\vspace{-0.5em}
\subsection{Clinical Sampling Bias and Observability Limits}

During our analysis, we noticed an interesting inverse relationship: the regions where our model performed best (like the precentral gyrus, $r=0.864$) had the smallest sample sizes ($N=2$), while the lowest performing regions (like the hippocampus, $r=0.197$) made up the bulk of the dataset ($N=124$). This disparity is not necessarily a flaw in the model; rather, it reflects how clinical SEEG sampling biases intersect with volume conduction physics.

When evaluating patients with focal epilepsy, the main goal is to find the seizure onset zone, which often originates in deep temporal structures. Because of this, neurosurgeons naturally implant most of the electrodes deep into the brain. On the other hand, the sensorimotor cortices are considered "eloquent cortex," so surgeons try to avoid placing electrodes there unless absolutely necessary, resulting in very sparse sampling for these superficial areas ($N \le 3$).

This creates a paradox for researchers using this data. Our model excels at reconstructing signals from superficial eloquent cortices because they are close to the scalp and readily observable. However, these regions make up only a tiny fraction of the dataset. Conversely, the deep structures that are heavily sampled are physically much harder to observe from the scalp \cite{he2026non,abdi2025eeg,nayak2004characteristics}. This sampling bias explains why about half of the ds004752 subjects (7/14) fell into our "non-viable" category; their electrode arrays were mostly targeting deep temporal structures, leaving very few contacts close to the surface.

\subsection{Practical Implications for Brain-Computer Interfaces}

Our findings regarding cross-subject generalization on the sensorimotor cortex could have practical implications for brain-computer interface (BCI) applications. In BCIs based on motor imagery, the precentral and postcentral gyri are exactly the areas we care about most \cite{vafaei2025transformers,mirchi2022decoding,dong2026bridging}. In our tests, these were also the regions where CAST achieved its highest reconstruction fidelity ($r > 0.7$).

This opens up the possibility for more accessible BCI systems. For example, a user could wear a standard scalp EEG cap, and a model like CAST could generate high-fidelity virtual iEEG signals from their motor cortex without surgical intervention or prolonged calibration. A pre-trained encoder could act as a universal feature extractor, and the system would only need a brief adaptation phase for the new user. Such deployment-oriented BCI systems are increasingly aligned with edge-AI and embedded real-time BCI platforms \cite{nguyen2025edge,nguyen2026edgessvep}.

More broadly, EEG-based machine learning has already been explored across SSVEP classification, decision prediction, multimodal neural decoding, object perception, and neurological screening tasks \cite{ou2023improving,tran2023early,tran2024multimodal,tran2024contrastive,tran2024eegssm}. Furthermore, in clinical settings like epilepsy surgery planning, this approach could potentially serve as a non-invasive screening tool. By estimating cortical iEEG activity beforehand, clinicians might be able to prioritize where to place electrodes during invasive monitoring, which could help reduce the surgical burden on the patient.

\subsection{Limitations and Future Directions}

We acknowledge several limitations in our study. First, while our depth gradient findings are statistically significant, the sample sizes for the very best-performing regions are small ($N = 2$ for the precentral gyrus, $N = 3$ for the postcentral gyrus). We view these as illustrative cases rather than population-level certainties. Second, our framework still relies on a short calibration phase (20\% of the test data). While this is a step forward from full within-subject training, achieving true "zero-shot" translation in future models would make clinical deployment even easier.

Third, our reconstruction accuracy for deep structures like the hippocampus ($r = 0.197$) and amygdala ($r = 0.233$) remains modest. Given that even advanced within-subject models using MRI data only reach $r \approx 0.3$--$0.5$ for these areas \cite{he2026non,dong2026bridging}, we suspect this is a fundamental physical limit rather than just a modeling issue. Lastly, because the GIN dataset lacked some anatomical labels (151 unlabeled channels), we could not perform a full region-by-region analysis on that cohort.

Moving forward, we suggest two strategies to improve cross-subject iEEG reconstruction for deep structures. First, adding geometric constraints from individual structural MRI scans could help the model account for differences in brain folding between people \cite{dong2026bridging}. Second, exploring probabilistic generative frameworks, like diffusion models, might help capture the high-frequency nature of deep neural signals better than standard regression models \cite{dong2025tf}.

\section{Conclusion}

In this study, we introduced CAST, a multimodal machine learning approach designed to reconstruct intracranial EEG from scalp recordings across different subjects. We tested our model on two independent datasets using a leave-one-subject-out method and observed three main outcomes.

First, we demonstrated that cross-subject transfer is possible for cortical contacts ($N = 313$) using a two-stage learning method. By training a universal encoder and using just a brief calibration phase (about 2 to 12 minutes) to adapt the decoder, the model successfully generalized to new patients. Under the best conditions, we saw peak correlations of $r = 0.864$ and $r = 0.732$ in sensorimotor areas.

Second, our results showed a statistically significant depth gradient ($p = 0.026$), where contacts closer to the surface were reconstructed more accurately than deep subcortical structures. This aligns with the physics of volume conduction and gives us confidence that the model is learning real neural relationships.

Third, by employing an observable channel selection strategy, we improved the model's performance on viable subjects to $r = 0.545$ ($R^2 = 0.333$) for the ds004752 dataset and $r = 0.458$ ($R^2 = 0.217$) for GIN.

Overall, our findings indicate that it is feasible to reconstruct cortical iEEG signals from scalp EEG for new subjects without needing extensive patient-specific training data. This methodology offers promising advancements for non-invasive BCI applications and clinical decision-making.

\bibliographystyle{named}
\bibliography{references}

\end{document}